\documentstyle[manuscript,prl,aps]{revtex}
 \begin{document}
 \title{On the Stability of the Compacton Solutions}
 \author{Bishwajyoti Dey$^1$, Avinash Khare$^{2,\dagger}$ 
\thanks{bdey@physics.unipune.ernet.in , $\dagger$ khare@iopb..stpbh.soft.net}} 
\address{1. Department of Physics, University of Pune, 
 Pune - 411007, India \\ and \\ 
2. Institute of Physics,
 Sachivalaya Marg, Bhubaneswar - 751005, India}
 
\maketitle
 \begin{abstract}
The stability of the recently discovered compacton solutions
is studied by means of both linear stability analysis as well
as Lyapunov stability criteria. 
From the results obtained it follows that,
unlike solitons, all the allowed compacton solutions are 
stable, as the stability condition is satisfied for arbitrary values of
the nonlinearity parameter. The results are shown to be true 
even for the higher order nonlinear dispersion equations for compactons. 
Some new conservation laws for the
higher order nonlinear dispersion equations are also presented.
\end{abstract}
\draft
 
\newpage 

The observed stationary and dynamical patterns in
 nature are usually finite in extent. However, most of the weakly
 nonlinear and linear dispersion equations so far studied admit solitary
 waves that are infinite in extent, although of a localised in
 nature. Therefore, the recently discovered compacton solutions (i.e.
 solitary waves
 with compact support), of the nonlinear dispersive $K(m,n)$ equations have
become very important from the point of study of the effect of
 nonlinear dispersion on pattren formation as well as the formation of
 nonlinear structures like liquid drops etc. 
The compacton speed depends on its height, but unlike the
 solitons, its width is independent of the speed, a fact, which seems to 
 play a very crucial role in its stability property. Compactons
 have remarkable soliton like property that they collide
elastically. However, unlike soliton collisions in an integrable systems, 
the point where two compactons collide is marked by the creation of low 
amplitude compacton-anticompacton pairs [1,2].
In fact, it is now known that
 the $K(m,n)$ system of equations are not integrable [1,2]. This suggest that
 the observed almost elastic collisions of the
 compactons is probably not due to the integrability and thus the mechanism
 responsible for the coherence and robustness of the compactons remains a
 mystery. Stability analysis of the compacton solutions may provide some
 clues in this direction. As has been said above, the stability 
of the compactons is crucial in the
 context of its applications in the study of pattern formation. Beside, the
 stability problem of the $K(m,n)$ equations is interesting because, for such
 equations with higher power of nonlinearity and nonlinear dispersion, the
 phenomena of collapse is possible. Also, in the context of soliton
 equations, from the stability analysis it has been shown that the higher
 order linear dispersion term stabilises the solitons [3,4]. In this regard
 it will be of interest to see what role does the higher order nonlinear
 dispersion term plays with respect to the stability of the compacton
 solutions of the$ K(m,n)$ type equations.
 
 In this communication we report on the stability analysis of the compacton
 solutions of the nonlinear dispersion $K(m,n)$ type equations as considered by
 Cooper et al [2]. We use both the linear stability
 analysis and the Lyapunov stability criteria to analyse the problem.
 We start with the $K(l,p)$ equations
 \begin {equation}
 u_t + u_x u^{l-2} +\alpha[2u_{3x}u^p + 4pu^{p-1}u_xu_{2x}
 +p(p-1)u^{p-2}u_x^3]=0
 \end{equation}
 These equations have the same terms as the $K(m,n)$ equations considered by
 Rosenau et al [1] but the relative weights of the terms are 
different, leading to the 
fact that, whereas the $K(l,p)$ equation (Eqn.(1)) can be derived 
from a Lagrangian, the 
$K(m,n)$ equation considered in [1] donot have a Lagrangian.
For the sake of comparison, it may be noted that the set of
parameters (m,n) in [1] corresponds to the set $(l-1, p+1)$ in Eqn.(1) [2]. 
 Assuming a solution to Eqn.(1) in the form of a travelling wave
$ u(x,t)=u(\xi)$, where $\xi=x-Dt$, Eqn.(1) reduces to
 the same  $K(l,p)$ equations considered by Cooper
 et al (Eqn.(7) in [2]) for the compactons solutions. The compacton solutions 
to Eqn.(1) are given by [2]
\begin{equation}
u(\xi) = [\frac {D}{2} (p+1)(p+2)]^{\frac {1}{p}} cos^{\frac {2}{p}} [\frac 
{p\xi }{2\sqrt {\alpha (p+1)(p+2)}}]
\end{equation}
in case $l=p+2$, $0<p\leq 2$ and for $\mid \xi \mid \leq \frac{\pi}{p}\sqrt{
\alpha (p+1)(p+2)}$ and $u(\xi)$ is zero otherwise. Note the width of the  
compacton is independent of its speed (amplitude). The fact that the width of 
the compacton will always be independent of its amplitude in case $l=p+2$ follows 
from the invariance of Eqn.(1) under the scaling transformation 
$x\rightarrow \alpha x, t\rightarrow \beta x, u\rightarrow \gamma x$. We 
now consider the stability of these compacton solutions.\\
{ \it Linear stability}- Eqn.(1) can be obtained 
 from the variational principle $\delta (H+DP)=0$.
 Using the relations 
 \begin{equation}
 I_n=\int_ {-\infty}^{+\infty}u^n dx \hspace{.2in}{\rm and} \hspace{.2in}
 J_2=\int_{-\infty}^{+\infty}u^pu_x^2 dx
 \end{equation}
we can write the corresponding conserved Hamiltonian and momentum as 
 \begin{equation}
 H_c=\alpha J_2 - \frac{I_{p+2}}{(p+1)(p+2)},
\hspace {.2in} P_c=\frac{1}{2} I_2
 \end{equation}
Using Eqn.(1) and the equation obtained from the scaling 
transformation $x\rightarrow \beta x$, we get
 \begin{equation}
 \frac{I_{p+2}}{(p+1)(p+2)}= \frac{(4+p)DP_c}{2(p+2)} 
\hspace{.2in} {\rm and}\hspace{.2in} \alpha J_2=
 \frac{pDP_c}{2(p+2)}
 \end{equation}
so that $H_c = -2DP_c/(p+2)$.
 Now considering the general scaling transformation 
$u\rightarrow \mu^{1/2}u(\lambda x)$, 
$H_c $ and $P_c$ are transformed to $H(\lambda ,\mu)$ and $P(\lambda ,\mu )$ and we
 get
 \begin{equation}
\Phi(\lambda, \mu) =\alpha\lambda\mu^{\frac{p+2}{2}}J_2 - 
\frac{\mu^{\frac{p+2}{2}}}{\lambda (p+1)(p+2)} I_{p+2} +
 \frac{\mu}{\lambda}DP_c
 \end{equation}
where $\Phi(\lambda, \mu)= H(\lambda, \mu) +DP(\lambda, \mu)$.
 The equations$ \frac {\partial \Phi}{\partial\lambda} =
 \frac{\partial\Phi}{\partial\mu}=0$ gives the stationary point at
 $\lambda=\mu=1$
 (compacton equation) and near this point, using the
 Taylor series we get for $\mu=\lambda$
 ( the transformation in that case does not change the
 momentum P) 
 \begin{equation}
 \delta\Phi(\lambda)=\delta H(\lambda) = (\lambda-1)^2 [
 \frac{\alpha(p+2)(p+4)J_2}{8} - \frac{p(p-2)(p+4)DP_c}{16(p+2)}]
 \end{equation}
 which has a definite sign. If it is positive (negative) the expression
 \begin{equation}
 H(\lambda)=\alpha\lambda^{\frac{p+4}{2}}J_2 -
 \frac{\lambda^{\frac{p}{2}}(p+4)DP_c}{2(p+2)}
 \end{equation}
 has a minimum (maximum) at $ \lambda = 1$.
 
 Now, let us assume that $u=u_c +v$, where $\mid v \mid <<1$ and the scalar product
$(u_c,v)=0$.
 Substituting this in Eqn.(1) we get after linearization
 \begin{equation}
 \partial_Tv = \partial_\xi \hat L v
 \end{equation}
 where$ \xi=x-Dt$ and T=t and the operator $\hat L$ is given by
 \begin{equation}
 \hat L=[D-u^{l-2}-2\alpha p u^{p-1}u_{2\xi} -2\alpha
 u^p\partial_{\xi}^2
 -\alpha p(p-1)u^{(p-2)} u_{\xi}^2 -2\alpha p u^{(p-1)}u_{\xi}
 \partial_{\xi}]
 \end{equation}
One can now run through the steps as given by Karpman [3] (see his Eqns. 
(29) to (32)) and show that as in his case, in our case too the sufficient 
condition for stability is that the scalar product $(\psi,  \hat L \psi ) >0$ 
where the operator $\hat L$ is given by Eqn.(10) while $\psi$ is a function 
in the subspace orthogonal to $u_c$. However, condition $(\psi, \hat L \psi) 
> 0$ is also associated with the extremum of $ H+DP$, since, using the 
relation $\delta(H+DP)=0$ one can show that the second variation of $H(u)$ and 
$P(u)$ at $u=u_c$ is given by
\begin{equation}
\delta^2(H+DP)_{u_c} = \frac{1}{2} \int _{-\infty}^{+\infty}(v,\hat L v) d\xi 
>0
\end{equation}
where the operator $\hat L$ is given by Eqn.(10). That is, if the condition 
$(\psi, \hat L \psi)>0$ is fulfilled, then $ H(u) + DP(u)$ has a 
minimum at $u=u_c$. Inversly, the minimum of $H(u) + DP(u)$ at $u=u_c$ is 
a sufficient condition of compacton stability with respect to 
small perturbation. Using Eqn.(8) we obtain the condition for the minimum 
of the perturbed Hamiltonian $H(\lambda )$ at $\lambda =1$ as 
$p+2 > p-2$ which is obviously true for any p. Thus we see that the 
condition for the Hamiltonian minimum (and hence the sufficient 
condition for the compacton stability) is satisfied for arbitrary values of the nonlinear 
parameter$ p$. This is unlike the soliton stability results, where it has been 
shown that the stability condition puts a restriction on the allowed 
values of the nonlinear parameter [3,4,6]. Note however that compacton 
solutions exist only for $p\leq 2$.

A la Karpman [3], one can show that the sufficient condition 
$(\psi, \hat L \psi )>0$ is also equivalent to the condition
 \begin{equation}
 (\frac{\partial P_c}{\partial D})>  0
 \end{equation}
From Eqns.(2) and (4) one can easily show that the sufficient
 condition for compacton stability (Eqn.(12)) is satisfied for arbitrary
 values of the nonlinearity parameter $p$. 
 It should be noted that this result is completely in contrast to
 the usual soliton stability results.
 It is not difficult to see why this is so for the compactons. From
 Eqn.(2) we see that the width of the compacton solutions is independent
 of its speed (amplitude) $D$, and the generic form of such compactons
is $u_c (\xi) = AD^bcos(c\xi )$, 
where the constants A,b and c depends on the
nonlinearity parameter. Hence $P_c =D^{2b} K, $ where K is D-independent. 
Therefore, $dP/dD > 0$ trivially since $b>0$. On the other hand, if
the width depends on speed (as in the case of solitons [3,4]) with the
generic form of the solution as $u(\xi) = AD^bcos(cD^a\xi)$, then 
$dP/dD > 0 $ only if $2b>a$ which will depend on the particular theory 
(soliton equations). It should be noted that the above stability 
condition (Eqn.(12)) is obtained by assuming that there is only one 
-ve eigenvalue for the operator $\hat L$ (Eqn.(10)). The 
validity of this conjecture has been proved from numerical experiments 
for many other systems, like the third and fifth order KdV equations as well 
as the nonlinear Schrodinger equations [3]. At present we donot have 
any evidence to show that this conjecture is also valid for our operator 
$\hat L$ (Eqn.(10)), except for the fact that the result that follows from 
using this conjecture also agrees with the result as obtained from the 
Hamiltonian minimum condition (Eqn.(11)) as well as an independent 
analysis of the stability by Lyapunov method as shown below.
 
These results are also true for the higher order
dispersion equations. For example, consider the
fifth order nonlinear dispersion equations, the $K(m,n,p)$ equation [5,7],
for
 the compactons. The Hamiltonian for such system is respectively 
given by [5]
 \begin{equation}
 H_c =\int_{-\infty}^{+\infty} [\delta \frac {u^{m+1}}{(m+1)} +\alpha 
u^{n-1}
 u_x^2 +\beta u^{p-3} u_x^4 +\gamma u^{p-1} u_{2x}^2]dx
 \end{equation}
and [7]
\begin{equation}
H_c =\int_{-\infty}^{+\infty} [\beta u^mu_x^2 -\alpha \frac{u^{p+2}}{(
p+1)(p+2)} -\frac{\gamma }{2} u^nu_x^lu_{2x}^2] dx
\end{equation}
 As has been shown in [5], the compacton solutions corresponding to 
the Hamiltonian in Eqn.(13) are allowed
 for the nonlinearity parameter $k=m=n=p$ in the range $2\leq k \leq
 5$ thereby meaning that the effect of the higher order nonlinear
 dispersion term is to increase the range of the nonlinearity parameter for
 which the compactons solutions are allowed. 
Considering the small perturbation $u=u_c +v$ as before, we can show that [8] 
even for these higher order nonlinear dispersion equations , the 
sufficient conditions for the compacton 
stability as given by Eqns.(11) and (12) 
is satisfied for arbitrary values of the 
nonlinear parameter 
$k$.\\
{\it Lyapunov stability}- The above theory of linear stability 
analysis for the compactons is based on the linearization of equations 
for compacton perturbations. 
This method has some inherent 
limitations connected 
with the linearization. So we present another approach  to the stability 
problem based on the Lyapunov method,  which, instead of linearization, 
uses sharp estimates. The effectiveness of this method has been 
demonostrated by Weinstein [9] and others [10]. In this method of 
analysis, it is sufficient
to prove that the Hamiltonian is bounded from below for fixed momentum $P$ 
and the compacton realizes the Hamiltonian minimum. We consider here the 
stability of the compacton solution of Eqn.(1). From Eqn.(3), we have
\begin{equation}
I_{p+2} \leq (max \ u^{\frac{p+4}{2}})^{\frac{2p}{p+4}} \int u^2 dx
\end{equation}
Also,
\begin{equation}
max \ (u^{\frac{p+4}{2}}) \leq \frac {p+4}{2} \int \mid u^{p/2} u_x \mid \mid 
u \mid dx
\end{equation}
Using Holder's inequality, we get
\begin{equation}
max \ (u^{\frac{p+4}{2}}) \leq \frac{p+4}{2} [\int u^pu_x^2 dx]^{1/2} [\int  
u^2 dx]^{1/2} \leq \frac{p+4}{2}J_2^{1/2} (2P)^{1/2}
\end{equation}
From Eqns. (4) we then have
\begin{equation}
H \geq \stackrel{\textstyle min}{J_2} [\alpha J_2 -\frac{1}{(p+1)(p+2)} 
(\frac{p+4}{2})^{\frac{2p}{p+4}} J_2^{\frac{p}{p+4}} (2P)^{\frac{2p+4}{p+4}}]
\end{equation}
Thus $H$ is bounded from below. On 
calculating the minimum of the right hand side we find 
 $H_{min}=-\frac{4}{p} \alpha J_2$
Thus we see that $H$ is bounded from below for arbitrary values of the 
nonlinearity parameter $p$. Now , from Eqns.(4) in Eqn.(5) we can 
immediately see that $H_c=H_{min}=-\frac{2DP_c}{(p+2)}$, i.e., 
the compacton realizes the Hamiltonian minimum
and hence  this proves the stability of the compacton solutions 
in the Lyapunov sense. Thus, the 
Lyapunov stability analysis show that all the allowed comacton 
solutions (i.e.  $p \leq 2$) are stable, since the condition 
for boundedness of the Hamiltonian and $H_c = H_{min}$ are valid for 
arbitrary values of the nonlinearity parameter $p$. It can be shown that 
[8] we also get similar result from the Luapunov stability analysis of the 
higher order nonlinear dispersion equation as given by the Hamiltonians 
in Eqns. (13) and (14).

Before concluding, we would like to mention that in our earlier paper [5] we had 
reported only one conservation law for the higher order $K(m,n,p)$ 
equations given by (Eqn.(2) in [5])
\begin{equation}
u_t +\beta_1 (u^m)_x +\beta_2 (u^n)_{3x} + \beta_3 (u^p)_{5x} =0 ,
\hspace{
.2in} m,n,p>1
\end{equation}
 We now find that, like the $K(m,n)$ equations  as considered
 by Rosenau et al [1], the higher order 
$K(m,n,p)$ equations also have four conservation laws for $m=n=p$, 
with the same conserved quantities as for the $K(m,n)$ equations, i.e. 
\begin{equation}
Q_1=u,\hspace{.1in} Q_2=u^{m+1},\hspace{.1in} Q_3=ucosx 
\hspace{.1in}{\rm and}\hspace{.1in} Q_4=usinx
\end{equation}
We have checked that even for the seventh order nonlinear disperison\\ 
$K(m,m,m,m)$ equation (see Eqn.(53) in [5]) there are four conservation 
laws as above. We suspect that even the generalised arbitrary 
odd-order nonlinear dispersion $K(m,m,m,m,....)$
equations (see Eqn. (36) in [5]) may also 
support similar four conservation laws. It should however be noted that for 
the $K(l,p)$ equations
(Eqn.(1) above) and its corresponding higher order $K(m,n,p)$ 
equation (Eqn.(13), (14) above), which are derivable from a Lagrangian and whose stability 
 problem is considered here, there are only three conservation 
laws [2,5].

To conclude,we would like to point out the important difference between 
the soliton and compacton solutions as obtained from the stability analysis 
of such solutions. Whereas the soliton solutions are allowed for arbitrary 
values of the nonlinear parameter, the stability condition on the soliton 
solutions puts restriction on the nonlinearity parameter for 
which stable soliton solutions are allowed [3,4,6,10]. On the other hand, 
the compacton solutions are allowed only within a certain range of the nonlinear 
parameter (the range is determined from the condition of the finite 
derivative of the compacton solutions at the edges [2,5]) and all the allowed 
compacton solutions (within this allowed range of the 
nonlinear parameter) are stable. Unlike soliton solitions, the 
stability of the compacton solutions does not put any additional constraint 
on the range of the nonlinear parameter. This result is true even for the 
higher order nonlinear dispersion equations for compactons, whereas 
for the soliton case the higher order linear dispersion term stabilizes 
the solitons with higher power of nonlinearity. It may be noted that we 
are unable to discuss the question of the stability of the compacton 
solutions as considered by Rosenau et al [1] since their compacton equations 
cannot be derived from a Lagrangian. We however suspect that their compacton 
solutions will also be stable. It would be nice if this can be shown in 
general.

\end{document}